\documentstyle[epsfig]{aipproc}

\begin{document}
\title{Status and Current Sensitivity of the CELESTE Experiment}

\author{M. de Naurois $^*$ for the CELESTE collaboration$^{\dagger}$}
\address{$^*$LPNHE, Ecole Polytechnique, Palaiseau 91128,\\
$^{\dagger}$CESR, Toulouse 31029, France,\\
CEN de Bordeaux-Gradignan 33175, France,\\
PPC, Coll\`{e}ge de France, Paris 75231, France,\\
LAL, Universit\'{e} Paris Sud, Orsay 91405, France,\\
GPF, Universit\'{e} de Perpignan 66000, France,\\
Charles University, Prague 18000, Czech Republic,\\
Joint Laboratory of Optics, Olomouc, Czech Republic.\\
}

\maketitle

\begin{abstract}
The CELESTE experiment uses the heliostats of an old solar farm in the French Pyrenees to detect gamma ray air showers by the atmospheric Cerenkov technique.
CELESTE has been operating since November 1999 with an array of 40 heliostats fully instrumented with 1GHz flash ADCs.
 Significant advances have been made in the detector simulations and in the data analysis techniques.
 We report here on results from recent observations of the Crab nebula above an energy threshold of 50GeV.
 The results and simulations illustrate the current sensitivity of the experiment.
\end{abstract}

\section*{Introduction}

CELESTE, described in detail in the experimental proposal \cite{proposal} and in \cite{NIM,berry,mdnthese}, uses the mirrors of a former solar electrical plant at the Themis site in the French Pyr\'en\'ees. We have been operating since October 1999 with forty, $54\ m^{2}$ heliostats which direct the \v{C}erenkov light, via secondary optics at the top of a 100~m tall tower, towards PMTs instrumented with flash ADCs. The ``standard candle'' of TeV astronomy, the Crab nebula, has been our primary target during this initial phase.

\section*{Simulation Results}

\begin{figure}[t!] 
\centerline{\epsfig{file=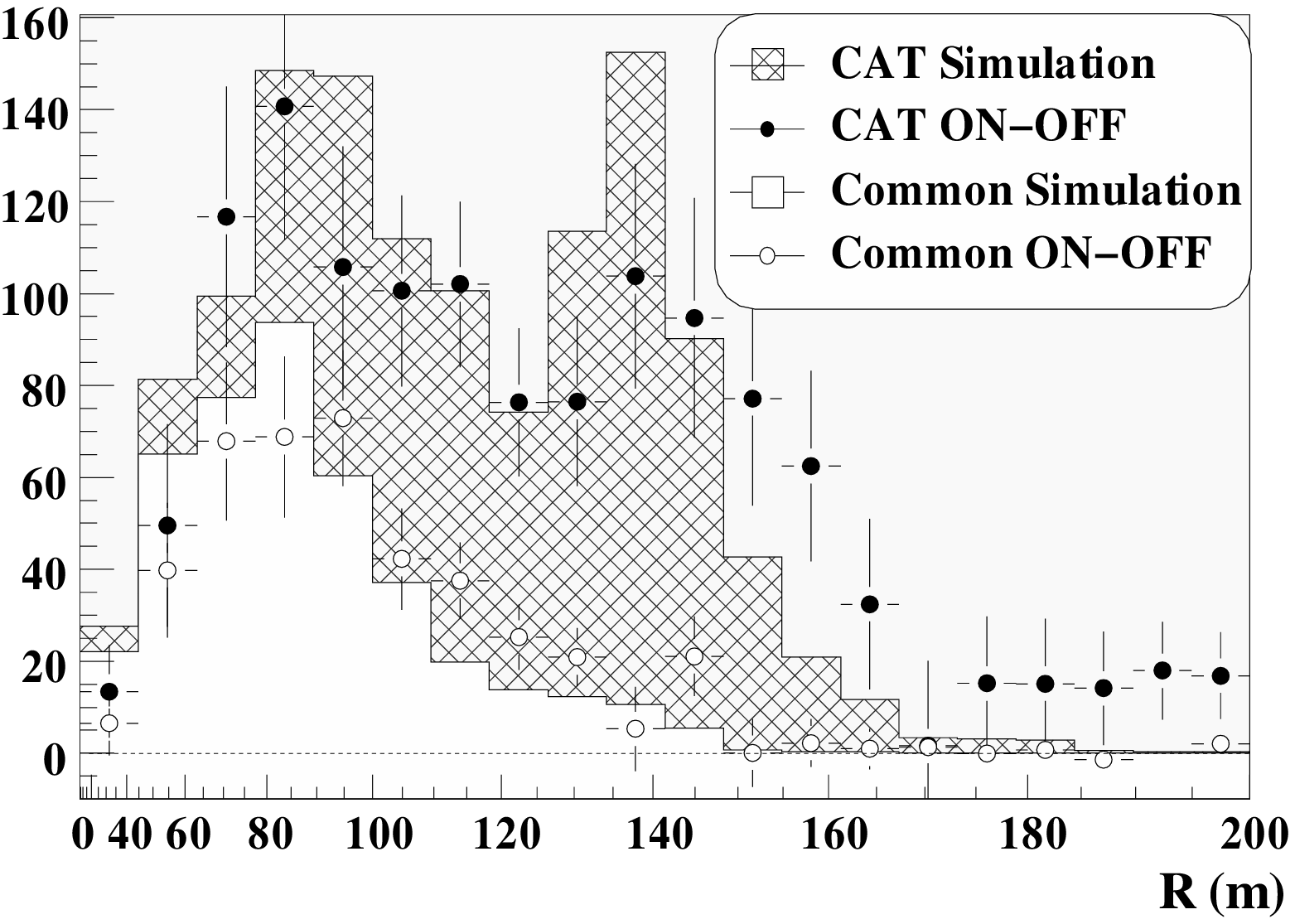,height=1.7in,width=3.0in}\epsfig{file=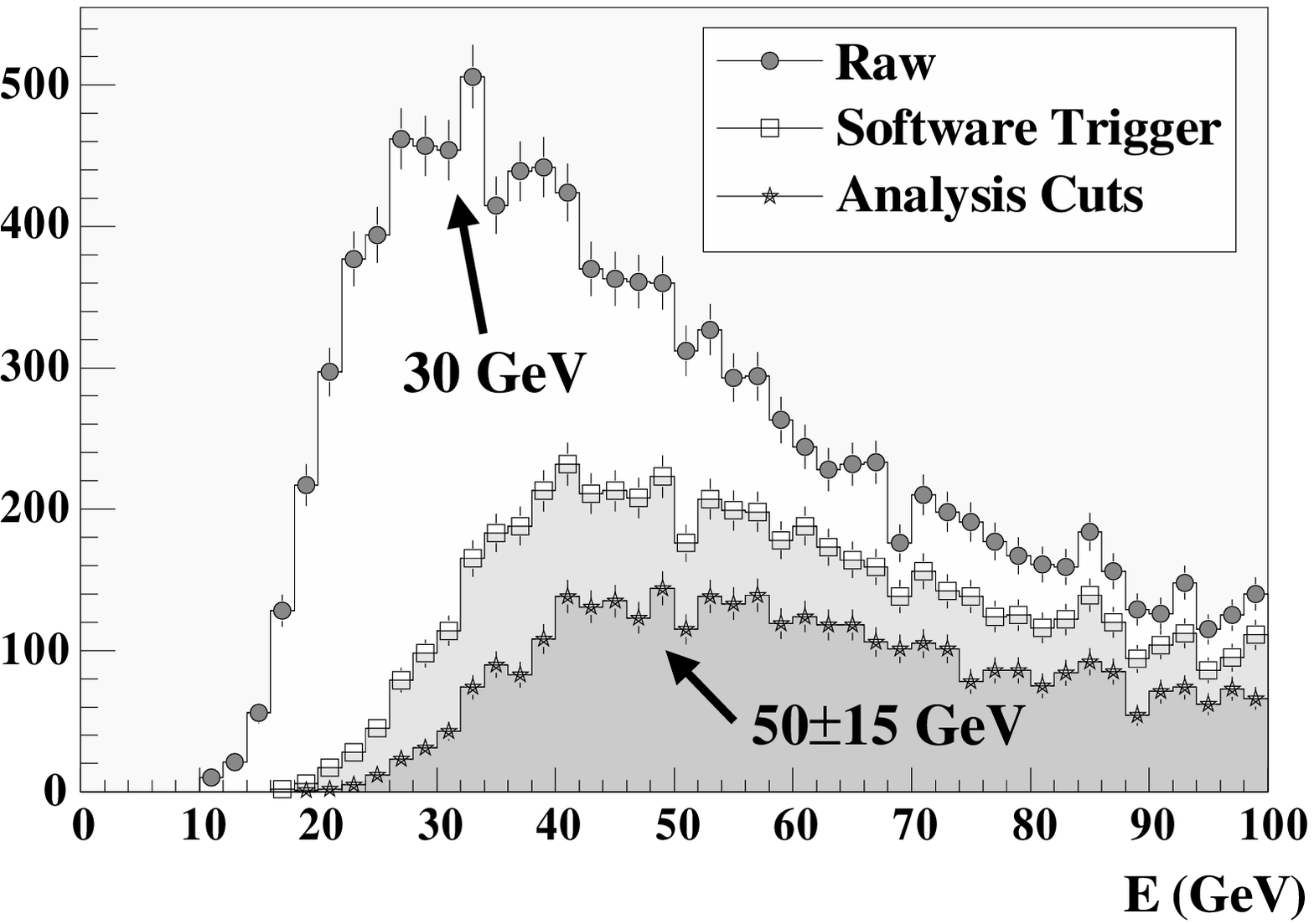,height=1.7in,width=3.0in}}
\vspace{10pt}
\caption{Simulation results. {\it Left:} The impact parameter (R) distribution as measured by CAT for real and simulated gamma rays. {\it Right:} The simulated response of CELESTE to an $E^{-2}$ power law gamma ray spectrum for a source at the position of the Crab at transit.}
\label{commonsurf}
\end{figure}

We have developed a simulation of CELESTE for use with standard air shower generators. 
 During the 1999-2000 Crab observing season we collected 13 hours of data with both CAT and CELESTE which can be used to partially check our simulations.
20\% of the CELESTE events, corresponding to $\sim$30\% of CAT events are common, as identified by their arrival time measured with GPS clocks at both experiments.
Analysis of the CAT dataset \cite{lebohec} results in an excess of 1268 gamma events over a background of 3131 hadrons.
 The same analysis applied only to the common events gives 418 gammas over 526 hadrons.
Thus, a CELESTE trigger doubles the signal to noise ratio in the CAT data, although it does not improve the significance as the data sample is smaller.
CAT measures the shower impact parameter with a resolution of $\sim$5~m.
Fig.~\ref{commonsurf} shows this reconstructed impact parameter for simulated data, and for excess events from the common Crab dataset.
The simulations reproduce the data, showing that the effective surface area for CELESTE in the energy region of the common CAT-CELESTE events is understood.

Also shown in Fig.~\ref{commonsurf} is the energy distribution of simulated events for an input $E^{-2}$ differential gamma ray spectrum.
At the raw trigger level the energy threshold, defined as the peak differential gamma ray flux, is $\sim$30~GeV.
We have found it necessary to apply a software trigger in order to remove the effects of night sky background differences between the ON source and OFF source control region, increasing our energy threshold to $\sim$50~GeV.
We also require $>10$ FADC peaks with an amplitude $>25$~digital counts to ensure the shower parameters are  measurable.

\begin{figure}[t!] 
\centerline{\epsfig{file=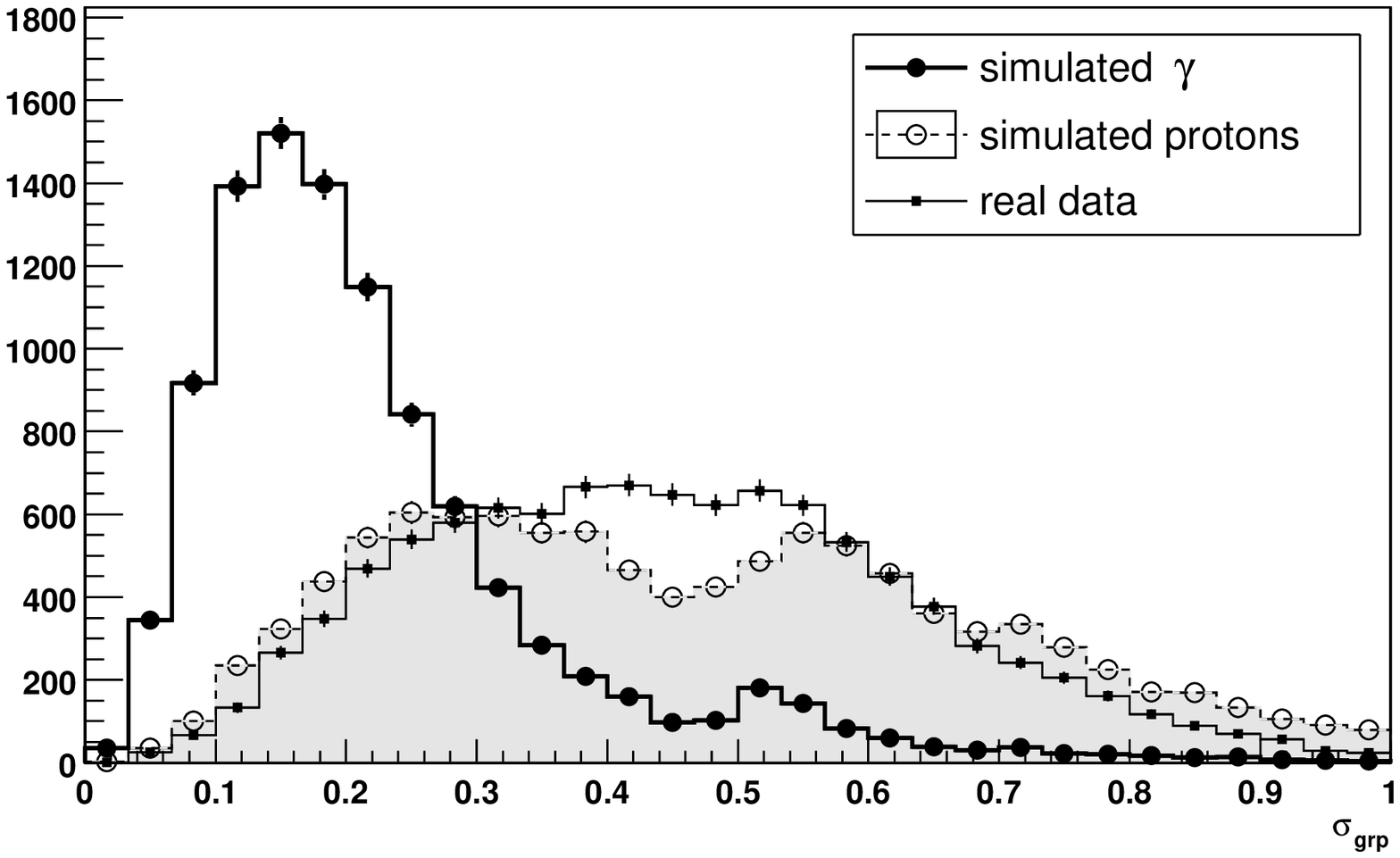,height=1.7in,width=3.0in}\epsfig{file=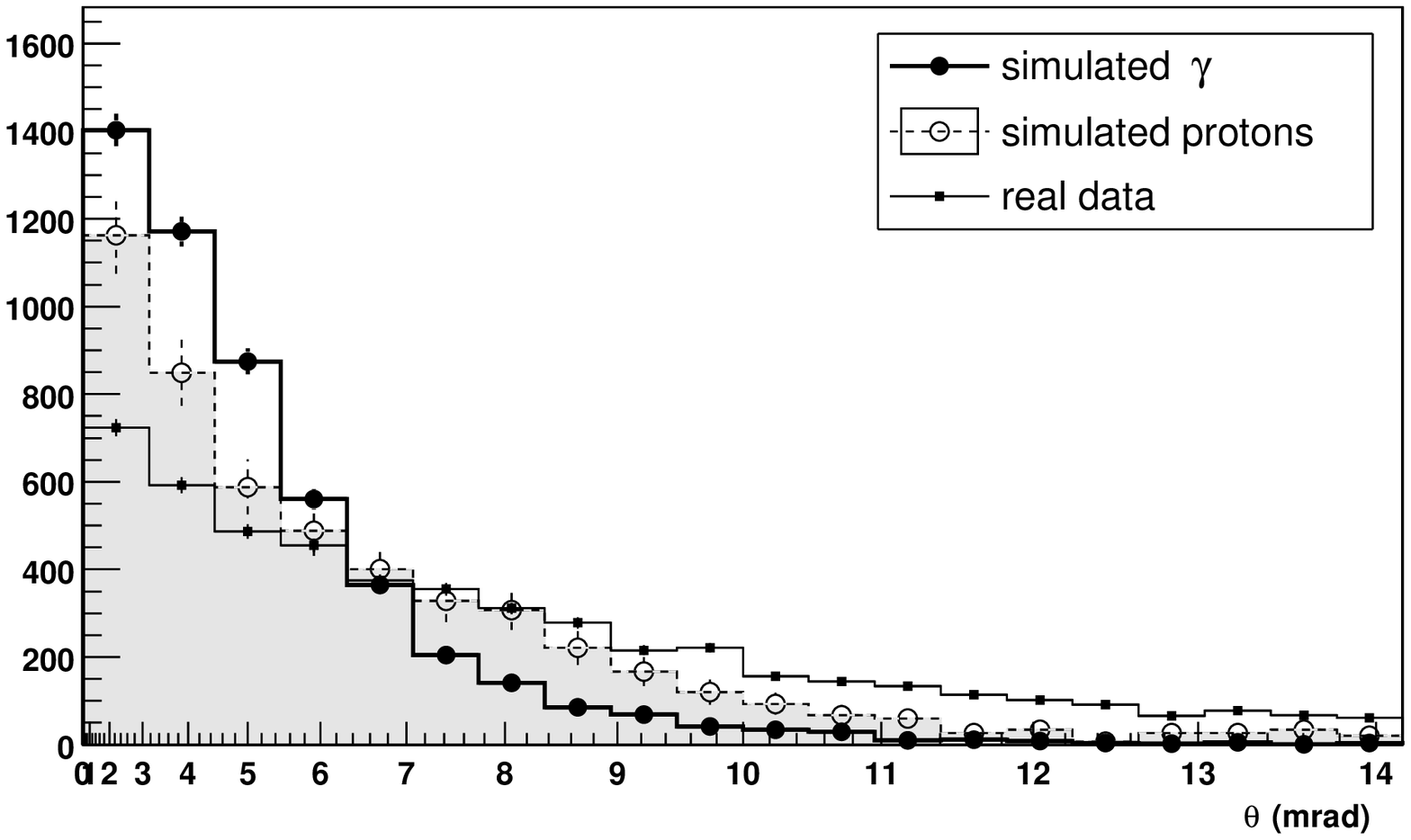,height=1.7in,width=3.0in}}
\vspace{10pt}
\caption{Analysis parameters. {\it Left:} The distribution of $\sigma_{grp}$ for simulated gammas, protons and real OFF source data. {\it Right:} The same for the shower axis angle, $\theta$.}
\label{params}
\end{figure}

We have used the simulations to develop our analysis techniques.
 The FADC peaks are fitted with a function modelled on the single photo-electron pulse shape and the reconstructed charge and timing information is used to derive the parameters of the shower.
 The homogeneity of the light distribution on the ground provides the best discrimination between hadronic and gamma ray showers.
 We define a parameter $\sigma_{grp}$, calculated by summing the FADC signals from 8 heliostats in 5 groups, corresponding to the hardware analogue sums used to trigger the experiment, then calculating the variance of the amplitude of the 5 pulses and normalising this variance to the mean amplitude.
 The distribution of $\sigma_{grp}$ for real data and for the simulations is shown in Fig.~\ref{params}. 
A cut at $\sigma_{grp}<$0.25 is expected to reject 80\% of the remaining hadronic background and retain 70\% of the gammas.

Also shown in Fig.~\ref{params} is the distribution of the angle, $\theta$, between the shower axis and the direction to the source reconstructed using two points:
the shower impact parameter at ground level, defined by the barycentre of the charge distribution over the heliostats, and the position of the shower core relative to the pointing direction at the mean altitude of maximum emission for gamma showers in the CELESTE energy range (11~km above the altitude of the experiment). The latter point is calculated by fitting a sphere to the measured arrival time of the \v{C}erenkov wavefront at each heliostat.
The small field of view of CELESTE (10~mrad fullwidth) reduces the observable differences between hadron and gamma showers; however, a cut at $\theta<$7~mrad is expected to reject $20\%$ of the remaining protons for $10\%$ of the gammas.
The effect of the analysis cuts on the energy distribution of the gamma showers is shown in Fig.~\ref{commonsurf}.

\section*{CELESTE observations of the Crab Nebula}
\begin{figure}[t!] 
\centerline{\epsfig{file=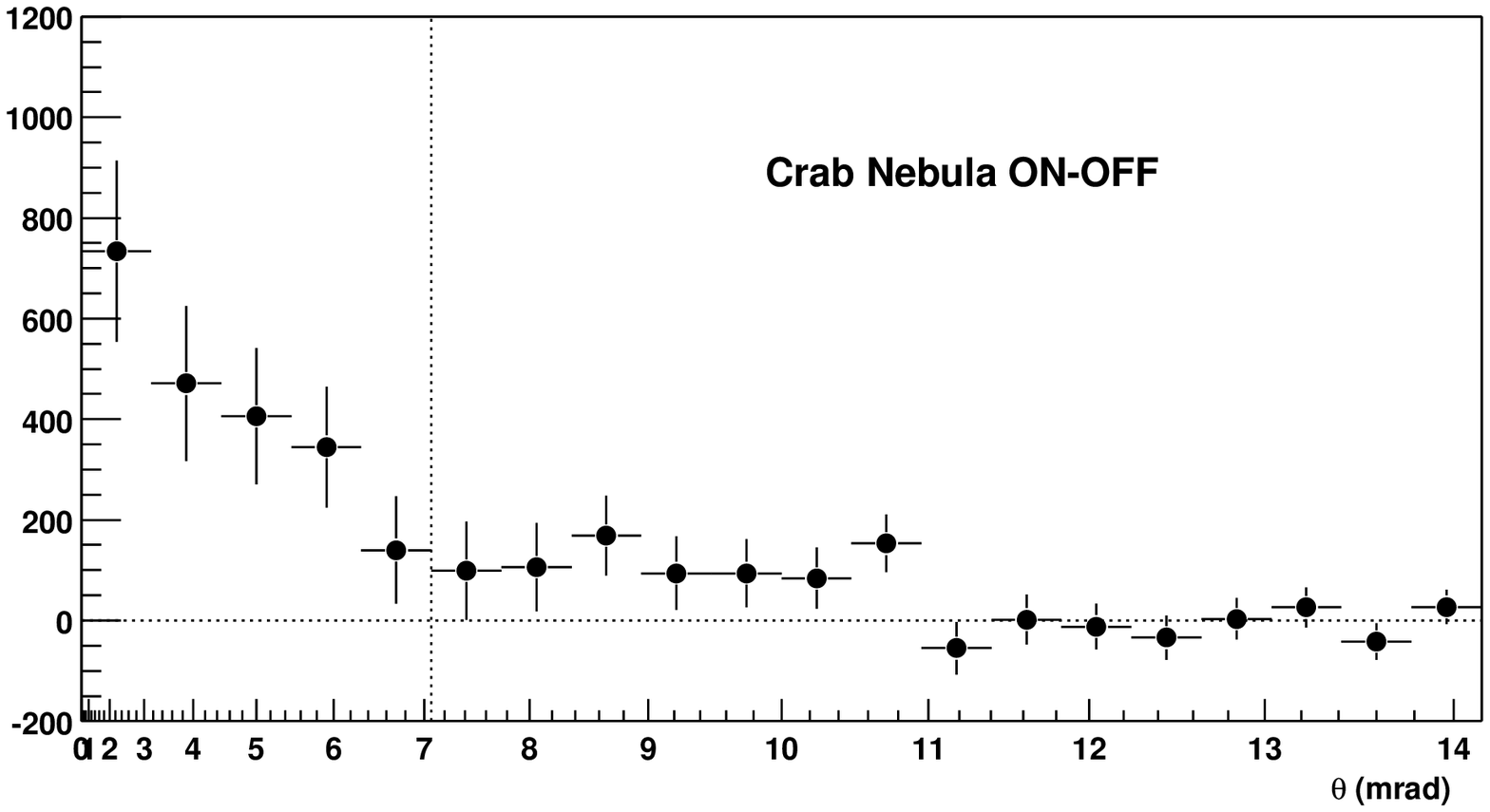,height=1.7in,width=3.0in}\epsfig{file=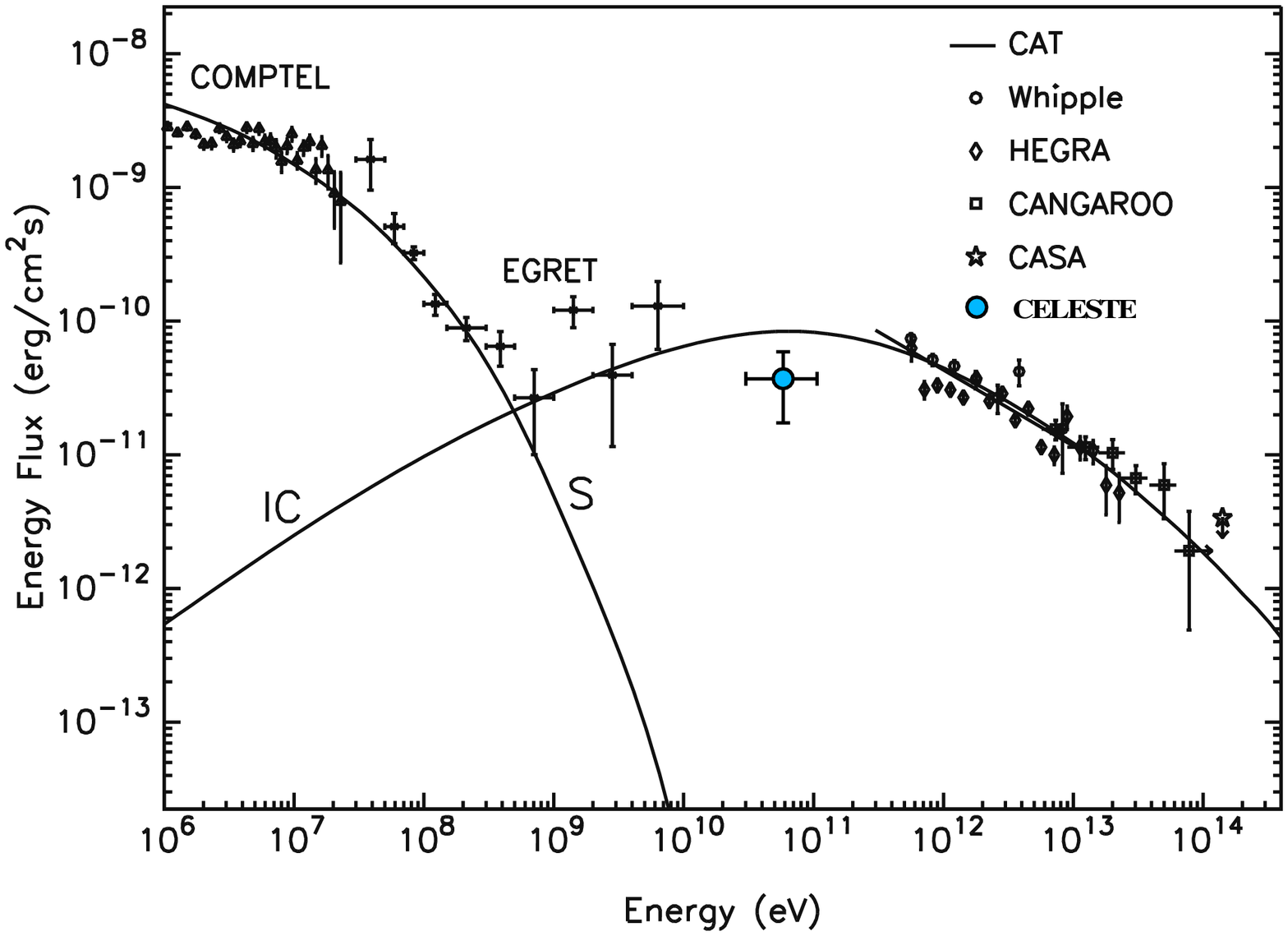,height=1.7in,width=3.0in}}
\vspace{10pt}
\caption[]{\label{theta}Results on the Crab. {\it Left:} The difference in the distribution of $\theta$ between ON and OFF source observations. {\it Right:} The Crab spectrum, including the preliminary CELESTE result and a model for the synchrotron and inverse Compton emission (after~\cite{aharonian}).}
\end{figure}

The Crab Nebula, as a strong steady source of VHE gamma rays \cite{wee89}, is the obvious first target for CELESTE.
We have collected 11.5h of ON source and corresponding OFF source data with all heliostats tracking a point 11~km above the altitude of the experiment.
Fig.~\ref{theta} shows the distribution of $\theta$ after all cuts, with an excess of events at small $\theta$ as expected.
Table~\ref{tcrab} shows the events remaining at each stage of the analysis.
The final significance of the result is 5.7~$\sigma$

\begin{table}[t!]
\caption{\label{tcrab} The number of events, excess, significance and signal to background ratio (S/B) after each cut for the Crab dataest.}
\begin{tabular}{l|ccccc}
 & ON & OFF & Excess & Significance & S/B \cr
\hline
Raw             & $850\,922$ & $845\,021$ & $5\,901$ & $4.5$ & $0.7\%$ \cr
Software Trigger       & $429\,533$ & $424\,881$ & $4\,652$ & $5.0$ & $1.1\%$ \cr
$N_{\mathrm{peaks}}\geq 10$   & $402\,452$ & $398\,242$ & $4\,210$ & $4.7$ & $1.1\%$ \cr
$\sigma_{\mathrm{grp}}<0.25$ & $84\,499$  & $81\,937$  & $2\,562$ & $6.2$ & $3.1\%$ \cr
$\theta\leq 7~\mathrm{mrad}$       & $49\,244$  & $47\,466$  & $1\,778$ & $5.7$ & $3.7\%$ \cr
\end{tabular}
\end{table}

We have calculated the normalisation factor to an assumed $\mathrm{E}^{-2}$ differential spectrum from the measured gamma ray rate, giving $ d\Phi / d\mathrm{E} = $
$$ (2.5 \pm 0.5_{stat} \pm 1.2_{syst})\times 10^{-8}/\mathrm{E}^2\ 
\mathrm{photons}\ \mathrm{cm}^{-2}\  \mathrm{GeV}^{-1}\  \mathrm{s}^{-1}$$
The CELESTE measurement is shown in Fig.~\ref{theta} along with the results from other experiments.
This is a preliminary estimate, as we are still quantifying systematic errors in various areas of the analysis, particularly in the comparison between the predicted and observed efficiency of analysis cuts.

We have also made some observations of the Crab using different tracking methods. A short (2h 10m) exposure using a double pointing method, wherein half of the heliostats point at 11~km altitude and the other half point at 25~km, appears to double our sensitivity by allowing us to sample a greater part of the shower.

\section*{Conclusions}
The CELESTE experiment has detected the Crab nebula with high significance (5.7~$\sigma$). Simulations have been used to develop our analysis procedure and also to estimate our energy threshold to be $\sim30$~GeV at the raw trigger level and $\sim50$~GeV after the analysis cuts.  
CELESTE will be upgrading to 54 heliostats shortly. This, in combination with improvements to the electronics and the double pointing strategy should greatly improve our sensitivity for the next Crab season.

\end{document}